# A Quantum Field Approach for Advancing Optical Coherence Tomography
## Part I: First Order Correlations, Single Photon Interference, And Quantum Noise

Mark E. Brezinski[1,2,3*]

[1]*Center for Optics and Modern Physics, Brigham and Women's Hospital, 75 Francis Street, Boston, M.A. 02115, USA*

[2]*Harvard Medical School, 25 Shattuck Street, Boston, M.A. 02115, USA*

[3] *Department of Electrical Engineering, Massachusetts Institute of Technology, 77 Massachusetts Avenue, Cambridge, M.A. 02139, USA*
[*]

Corresponding author: Mark E. Brezinski
T: 1-617-233-2802
F: 1-617-732-6705
Email: mebrezin@mit.edu

**Abstract**
Optical coherence tomography has become an important imaging technology in cardiology and ophthalmology, with other applications under investigations. Major advances in optical coherence tomography (OCT) imaging are likely to occur through a quantum field approach to the technology. In this paper, which is the first part in a series on the topic, the quantum basis of OCT first order correlations is expressed in terms of full field quantization. Specifically first order correlations are treated as the linear sum of single photon interferences along indistinguishable paths. Photons and the electromagnetic (EM) field are described in terms of quantum harmonic oscillators. While the author feels the study of quantum second order correlations will lead to greater paradigm shifts in the field, addressed in part II, advances from the study of quantum first order correlations are given. In particular, ranging errors are discussed (with remedies) from vacuum fluctuations through the detector port, photon counting errors, and position probability amplitude uncertainty. In addition, the principles of quantum field theory and first order correlations are needed for studying second order correlations in part II.

**Long Abstract**

Optical coherence tomography (OCT) is a micron scale ranging technology based on low coherence interferometry (LCI) that has found an important role in medical diagnostics. As OCT passes two decades of existence, the number of proven and potential applications continues to grow. But as expected, major advances in classical OCT technology seem to be slowing down. However, as in other fields of optics (and physics in general), utilizing the principles of quantum mechanics has the potential of producing paradigm-shifting advances in the technology. Almost all OCT theoretical work to date is classical, but a need exists for the advancement of OCT quantum mechanisms. This paper is the initial of a series on the quantum field analysis of OCT. It initially focuses on demonstrating some of the limitations of the classical treatment of OCT by examining first order correlations, primarily at the single photon limit, and closely related vacuum fluctuations. In this initial paper, concepts of treating the electromagnetic field (EM) field as a 'sea' of harmonic oscillators is reviewed (full quantization rather than semi-classical), as well as describing the basic mathematical tools for field quantization that includes annihilation/creation (and the related electric field operators) and density operators. Then first order correlations with OCT are modeled using single photon interferences dependent on indistinguishable paths. Classical linear interferometry results are yielded by the appropriate superposition of large numbers of single photon interferences. The paper concludes with a practical application. Here quantum noise sources are explored in OCT, primarily from vacuum fluctuation and photon count errors (PCE), which can be treated in the same context of first order correlations. These areas will be developed to advance OCT and identify areas where future work is needed. This includes accounting for polychromatic light resulting in more complex photon pressure effects, the fact backreflections are coming simultaneously from different depths, the differences in mass between the reference and sample arm (resulting in higher position probability uncertainty), and the target may behave at times like subsystems rather than a single unit. Second (and higher) order coherence, entanglement, and position probability amplitude uncertainty, among other topics, will be dealt with primarily in part II in the context of photon interactions. These areas are the focus of our group. However, to discuss these topics, the foundations in this paper are needed. By developing a quantum field approach to OCT, initially focused on single photon wavepacket interferences, we build a foundation for future OCT advances.

**OUTLINE**:
**1. Introduction**

**2. Optical Coherence Tomography**
    *2A. General Description*
    *2B. OCT Theory- Monochromatic Michelson Interferometry*

**3. Basic Concepts of the Quantized Optical Field**
    *3A. General Quantum Mechanical Principles*
    *3B. The Quantized Harmonic Oscillator*
    *3C. Fock States and the Quantum Harmonic Oscillator*
    *3D. Qualitative Analysis of Indistinguishable Paths*
    *3E. Single Photon Interference Along Indistinguishable Paths, the Basis of OCT*

**4. OCT Correlation Functions (Classical and Quantum Mechanical)**
    *4A. OCT Classical Correlation Functions*
    *4B. OCT Quantum Correlation Functions*
    *4C. Relationship Between the Quantum and Classical OCT Correlation Functions*

**5. Application of Quantum Optics to Advancing OCT: Quantum Noise Reduction**
    *5A. General*
    *5B. Quantum Noise Source: Qualitative*
        a. Position Probability Amplitude Uncertainty
        b. Vacuum Fluctuations at the Beam Splitter
        c. Photon Counting Error (PCE)
    *5C. Quantum Noise Source: Quantitative*
        a. General
        b. Vacuum Fluctuations at the Beam Splitter Causing Noise
        c. Combined Quantum Noise at the Detector
    *5D. Other Vacuum Fluctuations Errors*

**6. Conclusions**

**References**
**Figures**
**Appendix: Measurement and Decoherence**

1. Introduction

Optical coherence tomography (OCT) is a micron scale ranging technology based on low coherence interferometry (LCI) that has found an important role in medical diagnostics.  It is FDA approved and a clinical diagnostic in ophthalmology and cardiology.  And as OCT has passed its 20th year of existence, the number of potential applications continues to grow, with many new applications in clinical trials [1,2].  At the same time, paradigm shifts in classical OCT technology seem to be slowing down.  However, as in other fields of optics (and physics in general), quantum mechanics has the prospect of producing paradigm-shifting advances in the technology.  Unfortunately, quantum mechanics is often viewed primarily relevant to the microscopic world, so less effort goes into this area.  A surprising point of view coming from the optics field where solid-state quantum mechanics is so readily employed, for example, in optical sources and detection devices (i.e. light-matter interactions).  Example areas, as will be demonstrated, where understanding the quantum mechanics of first order correlations can offer OCT advances including reducing quantum noise and understanding position probability amplitude uncertainty.  But higher order correlations and interactions we believe offer even greater potential such as tissue characterization (ex: distinguishing lipid from nonlipid plaque) through the quantum properties of second order correlations, combined with varying position probability amplitude spreading, an area we have previously published [3-7].  The current article introduces a quantum field approach to OCT primarily by examining first order correlations and the ability to improve diagnostics.  First order correlations are treated in terms of single photon interferences through indistinguishable paths, with the field (and vacuum) being modeled by quantum harmonic oscillators.  In part II, second order correlations, entanglement, and related phenomenon (particularly position probability uncertainty) will be examined build from the theoretical framework of part I [6-8].

The paper begins by giving the classical analysis of OCT and then introducing concepts needed for analysis relevant to quantum field theory.  Then the quantum field approach to OCT is built initially on a quantized electromagnetic field (EM), the quantum harmonic oscillator, and single photon interferences.  Then classical and quantum mechanical OCT are compared quantitatively through the autocorrelation function.  Finally, how this becomes practical (for first order correlations) is demonstrated with quantum noise reduction and position probability amplitude uncertainty.  Quantum field theory of this type has its foundations in the work of pioneers such as Dirac, Feynman, Caves, Loudon, Glauber, Ben-Aryeh, Teich, Mandel, and Saleh (among others).  We utilize this work from other fields to build on the quantum field analysis of OCT.

## 2. Optical Coherence Tomography
*2A. General Description*

The classical principles behind OCT are described in detail elsewhere and so will only be discussed here superficially [1}.  We begin with the initial discussions of the classical theory of interference and then compare it to the quantum theory through the autocorrelation function.  This will begin illustrating the limitations of classical descriptions of OCT.

With respect to OCT, in the 'classical embodiment', low coherent light from a source is directed at a beam splitter.  Half the light is directed at a reference arm (which contains a distal mirror) and half at the sample arm (figure 3).  The optical path length in the reference arm is being changed continuously. Light reflects at the distal end of both arms. When light recombines at the beam splitter, interference occurs when the delay in both arms is within the coherence length. This is the classical explanation but would not account for, for example, single photon interference (which is the basis of all first order linear interferometry) as we will see in the next section.

*2B. OCT Theory- Monochromatic Michelson Interferometry*

OCT systems are generally based on Michelson interferometry. So we will discuss classical interferometry and OCT for a Michelson interferometer. We will start initially using monochromatic light and then polychromatic Gaussian light. The reason for including the monochromatic classical derivation is that when we develop quantum first order correlations through single photon interference, we will start with monochromatic light. Comparisons will be made between quantum and classical mechanics.

In a Michelson's interferometer, light from the source, expressed in terms of the electric field $E_{so}$, is directed at the beam splitter. The beam splitter 'divides the light' (this is a classical description) into $E_r$ and $E_s$, where r is the reference arm and s is the sample arm respectively (directed at perfect mirrors). These two complex monochromatic plane waves have the same frequency, wavenumber, and phase (we are ignoring the phase shifts from the splitter) as they are split. We will ignore any losses occurring within the interferometer itself from scattering or absorption. We will also assume reflectivity off both mirrors is 100%. After reflecting off the two mirrors, the light recombines at the beam splitter, so that the electric field at the detector is $E_D = 1/\sqrt{2} E_r + 1/\sqrt{2} E_s$. We will be examining what happens with monochromatic light when the two arms have different path lengths before combining at the beam splitter. Since $k = \omega/c$, we will represent these waves as:

$$E_r(x) = \frac{1}{\sqrt{2}} E_{S0} e^{i(\omega x_r)/c}$$

$$E_S(x) = \frac{1}{\sqrt{2}} E_{S0} e^{i(\omega x_S)/c} \quad (1)$$

The only difference between the two waves is the distance they have traveled. So $E_D$ equals:

$$E_D(x) = \frac{1}{\sqrt{2}} E_r(x) + \frac{1}{\sqrt{2}} E_S(x+a) = \frac{1}{\sqrt{2}} E_{S0} e^{i(\omega x_r)/c} + \frac{1}{\sqrt{2}} E_{S0} e^{i(\omega x_S + a)/c} \quad (2)$$

Now what is generally measured at the detector is irradiance and not the electric field, which is the time average of the square of the electric field. This is represented by $I = \varepsilon v \langle EE^* \rangle_T$. The T stands for time average. This translates to (ignoring the constants $\varepsilon v$):

$$I_D(x) \propto \langle E_D E_D^* \rangle_T = \frac{1}{2} \langle E_r E_r^* \rangle_T + \frac{1}{2} \langle E_S E_S^* \rangle_T + \frac{1}{2} \langle E_r^* E_S \rangle_T + \frac{1}{2} \langle E_S^* E_r \rangle_T \quad (3)$$

$E_r^*$ and $E_s^*$ are the complex conjugates of the electric field in the reference and sample arm respectively. This becomes:

$$I_D(x) \propto \langle E_D E_D^* \rangle_T = \frac{1}{2} I_r(x) + \frac{1}{2} I_s(x) + \frac{1}{4} \langle E_{S0}^* e^{-i(\omega x_r)/c} E_{S0} e^{i(\omega x_S)/c} + E_{S0} e^{i(\omega x_r)/c} E_{S0}^* e^{-i(\omega x_S)/c} \rangle_T$$

(4)

In the right side of equation (4), the first two items represent the DC irradiance from simple and reference arms (which does not carry ranging information) respectively as

$$I_r(x) \propto \langle E_r E_r^* \rangle_T \quad \text{(5-a)}$$
$$I_s(x) \propto \langle E_s E_s^* \rangle_T \quad \text{(5-b)}$$

We want equation (4) to be in the form:

$$I_D(x) = \frac{1}{2} I_r(x) + \frac{1}{2} I_s(x) + \left\langle \text{Re}(\frac{1}{2} E_{S0} E_{S0}^* e^{i(\omega x_r/c - \omega x_S/c)}) \right\rangle_T \tag{6}$$

To achieve this, we will use an identity:

$$\cos\theta = \frac{e^{i\theta} + e^{-i\theta}}{2} \quad \text{Or} \quad 2\cos\theta = e^{i\theta} + e^{-i\theta} \tag{7}$$

Equation 4 now has the form:

$$I_D(x) = 1/2 I_r(x) + 1/2 I_s(x) + 1/4 \langle E_r E_s^* \rangle_T \cos\Theta \tag{8}$$

As a reminder, $E_r$ and $E_s$ are complex quantities. The terms $I_r(x)$ and $I_s(x)$ are DC terms of regular rapidly oscillating electrical fields from the source, but the third term is the interference term. When $\Theta$ is 0 or a multiple of $\pm 2\pi$, the cosines value is maximum at 1. The value of the interference term is then $1/4 \langle E_r E_s^* \rangle_T$. This is total constructive interference and the waves are said to be in phase (ignoring polarization effects). When the value of $\Theta$ is a multiple of $\pm \pi$, the value of the cosine is –1 and $I_D(x)$ is at a minimum. This situation is called total destructive interference. From equation (5-a) and (5-b), it is known that $I_r(x) = I_s(x) = 1/2 \langle E_{so} E_{so}^* \rangle_T = 1/2 I_{so}$ (when reflectivity is equal in both arms). Equation 7 becomes:

$$I_D = 1/2\ I_{so} + 1/2\ I_{so} \cos\Theta = 1/2\ I_{so}\ (1 + \cos\Theta) \tag{9}$$

When $\Theta$ equals is zero, the intensity at the detector becomes $I_{so}$. When $\Theta$ is $\pm\pi$, the intensity in the detector arm is zero and by conservation of energy, all the intensity is in the source arm. This derivation implies that interference is occurring between both arms of the interferometer. Essentially the same results will be seen with large photon numbers via a quantum field approach, but substantially different results occur at low photon numbers. The concept that 'light from both arms interferes' does not hold when derived from quantum mechanics as will be seen.

## 3. Basic Concepts of the Quantized Optical Field

*3A. General Quantum Mechanical Principles*

In the OCT community, the quantum field approach to EM fields is not commonly addressed. So in the next several sections, the mathematical principles will be provided from quantum field theory needed for a full quantization approach to OCT. For those with a command of quantum field theory, this section may be rudimentary. It should be noted that in optics, a full quantization approach is not always necessary and can be reasonably approximated by treating the system semi-classically. In the semi-classical approach, matter is described quantum mechanically (ex: quantized photon absorption of the detector) but the EM is treated classically plus the addition of vacuum fluctuations [10-12]. But many phenomena relevant to our discussion of advancing OCT including vacuum fluctuations in a beam splitter, two photon interferometry, and entanglement can not be reasonably described semi-classically [13,14]. These require full or second order field quantization (i.e. the field needs to be quantized). When second order quantization is usually utilized, with origins in the work of Dirac, again the field is approximated as a bath of quantized harmonic oscillators. The vacuum represents the lowest energy level of the oscillators (which is non-zero) [15]. Then energy added to the system occurs in increments of EM quanta or photons (hω). This is among the concepts discussed.

We will be using quantum formalism throughout the text. In this quantum formalism, real values of observables (energy, position, momentum...) are represented by Hermititian operators (^ carrot signifies operators) and the initial/final states within kets/bras (| > or < |), respectively. Wavefunctions are used to represent the state itself in most elementary quantum texts. A wave function or wavefunction is a probability amplitude in quantum mechanics describing the quantum state of a particle (or system of few particles). It is almost always a pure state. But wavefunctions have substantial limitations for use in this paper (though they will be used in several incidences for convention), such as not easily representing mixtures, particularly when coherences are involved. So in the majority of this text density operators, symbolized by ρ, are used for state representation. Many texts treat quantum density operators as analogous to classical statistical matrices, which is an inappropriate interpretation [16,17]. The density operator is an operator acting on Hilbert space whose representation includes non-classical coherences. But it does not represent a priore distribution itself. But the trace of the density operator can produce observable averages (such as particle or energy distributions).

*3B. The Quantized Harmonic Oscillator*

Quantized harmonic oscillators will represent the field. Here, first we must establish what base states we are going to work in. There are various base states for quantizing the EM field, such as with position/momentum or Glauber's coherent states. But a number state representation (Fock state) is a particularly useful basis as will be evident momentarily [6-7, 16-17]. Here, the quantum harmonic oscillators that make up the field go up and down levels by the value of photons. These are formally added or subtracted using annihilation and creation operators (that will be shown to typically have a linear relationship with the electric field operators). Therefore, state change is performed by very simple algebra making this basis advantageous. This is as opposed to working with, for example, position and momentum operators found in introductory quantum mechanics texts that are challenging to work with.

As stated, the field (which can be just the vacuum or the vacuum occupied by photons) will be described in terms of a bath of quantized harmonic oscillators. As has been previously performed, second quantization will be derived beginning from the Hamiltonian operator of a classical harmonic oscillator and then extending this as being analogous in form to Schrödinger's time independent equation (STIE). We are pursing the Hamiltonian to generate the quantized energy levels of the vacuum/field and express it in terms of the creation operators. There are more extensive derivations then the Hamiltonian using a Lagrangian and Maxwell's equations that would provide even further insight (and will be the source of discussion in future work), but derivations of this type can be found elsewhere and the Hamiltonian is sufficient for purposes here [18].

The classical Hamiltonian operator (one of several equivalent forms) for a classical, non-quantized harmonic oscillator is given by the kinetic and potential energy (here one dimension):

$$\hat{H} = \frac{\hat{p}^2}{2m} + \frac{1}{2} m\omega^2 \hat{x}^2 \qquad (10)$$

where again the carrot means that this is in operator format. The first term containing momentum is the kinetic energy term and the second the potential energy. To put it in the form of STIE, momentum will be represented in its quantum mechanical equivalent $\hat{p} = -i\hbar \frac{\partial}{\partial x}$. The equation then becomes STIE (one dimension):

$$E\Psi(\mathbf{r}) = \frac{-\hbar^2}{2m} \nabla^2 \Psi(\mathbf{r}) + V(\mathbf{r})\Psi(\mathbf{r}) = \hat{H}\Psi$$

$$V(x) = \frac{1}{2} kx^2 = \frac{1}{2} m\omega^2 x^2$$

$$\frac{-\hbar^2}{2m} \frac{d^2\Psi(x)}{dx^2} + \frac{1}{2} m\omega^2 x^2 \Psi(x) = E\Psi(x)$$

(11)

The equation is time independent with respect to observables (a constant of energy). So the classical Hamiltonian is analogous to STIE. The first term on the left of the equal sign (last equation) is the quantum analogy of the classical harmonic oscillator of momentum or mass times acceleration. The second term on the left of the equal sign is the potential, the equivalent of the spring constant times the displacement. The potential is represented by mass times the angular momentum squared ($\omega^2$) times the displacement operator squared. We will focus on the quantized energy solutions (and not the spatial solutions).

Three points about this derivation that is not representative of the remainder of the paper. First, the wave function is used here rather than a density operator staying consistent with convention for STIE presentation. But through the majority of the paper the density operator will be used to represent the state rather than the wave function, which accounts for complex mixed states. Second, we were working in the basis states of position and momentum, which we stated is challenging to use with a quantized optical field. This again is convention, only for the purpose for comparison with the common form of the classic harmonic oscillator, and we will be using Fock states through the majority of the paper. Third, we used the differential form but predominately we will be using Heisenberg's matrix approach through the rest of the paper.

The position probability solutions for a quantum harmonic oscillator are not the point of interest in this paper. Instead, the energy states are the focus because this is what the field will be quantified in terms of.

The lowest energy state of the harmonic oscillator, from equation 11, is the vacuum that has non-zero energy (due to the uncertainty principle), where the energy of each angular frequency of the vacuum is given by:

$$E_k = \frac{1}{2}\hbar\omega_k \quad (12)$$

and the expectation value of the vacuum is given by the sum over frequencies (where k values included are controlled primarily by the volume):

$$\langle E \rangle = \frac{1}{2}\sum_k E_k \quad (13)$$

It will be discussed below that the frequencies allowed will largely be a function of the volume. The addition of photons (the EM field) to this vacuum will be dealt with in the next sections. Both the quantized vacuum and photons will be important in understanding the quantum properties for advancing OCT application.

*3C. Fock States and the Quantum Harmonic Oscillator*

We need base states for our quantum harmonic oscillator. As stated, position and momentum are difficult base states to work in. The coherent states representation made famous by Glauber would be an improvement and is used by many authors, but it involves operators that are not Hermititian and a coherent state basis that is not orthogonal [16,17]. A vastly simpler approach is the use of quantum number states (n) or Fock numbers. The number states can be viewed as equally spaced energy levels of the harmonic oscillator. The sequential energy levels of the oscillator are separated by the energy of a photon (hω).

Using the concept that the state energy increases in increments of photons, with definite photon number n (monochromatic representation), is given by:

$$E_{n_k} = \hbar\omega_{n_k}(n_i + \frac{1}{2}) \quad (14)$$

This is the energy solution to Schrödinger's equation (equation 11) for a monochromatic quantum harmonic oscillator. Expressed in the operator form for obtaining real values of states, the numerical values are replaced by $\hat{E}$ and $\hat{n}$, described in more detail below. We will be using these operators throughout the remainder of the paper to generate observables. Here n is the photon number for a monochromatic state of a definite number of photons. The value of n is zero for the vacuum state at a given wavenumber and n is positive in the presence of an EM field.

But as with other observables in quantum mechanics, the number state can be in a superposition, so we can not simply represent the number state by a specific number of photons but rather need a state vector, $|n\rangle$ and a number operator $\hat{n}$. The eigenvalues (n) are then given by:

$$\hat{n}|n_i\rangle = n|n_i\rangle \quad (15)$$

Here $|n_i\rangle$ is the eigenstate that yields an eigenvalue n. Superposition of these eigenstates can occur analogous to, for example, spin states where the corresponding eigenstates are the Pauli spin states and operators are the Pauli spin operators. The number operator is also often referred to as the ladder operator (the energy levels are equally spaced like a ladder in this single oscillator example), separated by the value of a photon.

The derivation can be found in most elementary quantum texts but the number operator can be expressed in terms of the annihilation and creation operators (which we will use extensively):

$$\hat{n} = \hat{a}^+ \hat{a} \qquad (16)$$

These operators work on number states such to increase or decrease the energy and as well as photon numbers (monochromatic).

$$\hat{a}|n_i\rangle = \sqrt{n}|n_{i-1}\rangle$$

$$\hat{a}^+|n_i\rangle = \sqrt{n+1}|n_{i+1}\rangle \qquad (17)$$

The first is the annihilation operator while the second is the creation operator. We are working in the basis of photon number so a number state (at a given wavenumber) can be built up from the vacuum using the creation operator:

$$|n\rangle = \frac{1}{\sqrt{n!}}(a^+)^n|0\rangle \qquad (18)$$

It should be clear that we could build any photon number (at a given frequency) up in this manner and that different numbers for different frequencies. But using the annihilation operator on the vacuum has no physical significance.

$$a|0\rangle = 0 = \langle 0|a^+ \qquad (19)$$

It should also be noted we could always convert this to the position-momentum operator basis if necessary, where transforms exist between the two:

$$\hat{a} = \sqrt{\frac{m\omega}{2\hbar}}\left(\hat{x} + \frac{i}{m\omega}\hat{p}\right)$$

$$\hat{a}^+ = \sqrt{\frac{m\omega}{2\hbar}}\left(\hat{x} - \frac{i}{m\omega}\hat{p}\right) \qquad (20)$$

But it can be seen that, in the momentum-position basis, but it is no longer a simple matter of photon addition or subtraction.

The objective now is to express the quantum harmonic oscillator Hamiltonian in terms of the Fock states. So $\hat{n}$, $\hat{a}$, and $\hat{a}^+$ will play an important role in our derivation of the field in terms of the quantum harmonic oscillator. The Hamiltonian for a single frequency, in terms of Fock states rather than position-momentum, is given by:

$$\hat{H}|n\rangle = \hbar\omega\left(\hat{a}^+\hat{a} + \frac{1}{2}\right)|n\rangle = \hbar\omega\left(\hat{n} + \frac{1}{2}\right)|n\rangle \qquad (21)$$

If n is a given number of photons of the same frequency, the Hamiltonian gives the energy operator. When looking at the harmonic oscillators containing frequencies over a finite range, the Hamiltonian becomes:

$$\hat{H}|\{n_k\}\rangle = \left(\sum \hbar\omega\left(n_{k_l} + \frac{1}{2}\right)\right)|\{n_k\}\rangle \tag{22}$$

So the Hamiltonian is defined from the sum of the number states at each wavenumber. This gives a description of the state of the field in terms of harmonic oscillators and the quantized photon occupation at each frequency, the quantized EM field. This section quantizes the vacuum and EM field using modeling with the quantum harmonic oscillator. We will extend this concept later to represent the low coherence EM used in OCT, where we will be primarily focused on the state in terms of a photon numbers and the energy given by the electric field operators.

In much of the text, monochromatic single photons will be used, particularly with single photon interference. Larger number of photons will be used when we introduce coherence functions (both quantum and classical). The monochromatic single photon wavepacket (ignoring polarization) is define here:

$$|1_\omega\rangle \equiv \hat{a}^+(\omega)|0\rangle \tag{23}$$

The 1 represents a single photon state. Obviously, the notion of a localized monochromatic wave packet has no classical analogy. This is because classically, a monochromatic wave would be infinite in extent rather than a packet.

*3D. Qualitative Analysis of Indistinguishable Paths*

The classical description of OCT or LCI is expressed in terms of superposition of wavefronts. But this does not explain, for example, single photon (or particle) interference or account for vacuum fluctuations, the latter being an important noise source. Vacuum fluctuations were accounted for in part in the description of the field as quantum harmonic oscillations in the previous section. In describing quantum OCT, we need to derive interference in terms of single photon events (because single photon interference is occurring) rather than superposition of wavefronts. Interferometry can be performed one photon at a time, which is not accounted for by the wavefront superposition description. Critical to this analysis will be the fundamental relationship of path indistinguishability and interference, which will be dealt with here qualitatively and quantitatively in subsequent sections.

It will also be demonstrated that for first order coherence, interference is generated when single particles/photons have indistinguishable potential paths. Path indistinguishability is at the heart of all linear interferometers. The importance of path indistinguishability will first be illustrated qualitatively with a Young's interferometer, as most are familiar with this experimental set-up from introductory physics courses (figure 1). The same analysis holds with a Michelson interferometer (OCT) as will be seen, but the Young's interferometer used initially simplifies the initial derivation (including eliminating issues associated with the beam splitters discussed below). We will then take the results and apply them to OCT.

The Young's interferometer has a barrier of two slits that limits the action to two indistinguishable paths. We are ignoring the path integral of all potential paths that is not needed to demonstrate the principles. In this Young's experiment, the source will be a neutron beam (rather than photons) entering the interferometer one at a time [22]. Neutrons are used instead of photons because it is easier demonstrating the influence of environmental interactions using collisions of particle with mass. We will not initially use massless photons as they are less susceptible to these environmental interactions. We could use photons but the design would need to be more elaborate. These environmental interactions will be used to demonstrate the relationship between path indistinguishability/distinguishability and interference. Also, the use of particles generally considered 'solid' further emphasizes the interference observed should not be considered the superposition of classical waves as described in the previous classical OCT section.

So in the classical description of Young's experiment (figure 1), if one or the other slit is blocked, the neutrons are registered on the screen with no interference pattern (NI). If both slits are opened, it is easy to appreciate if we were dealing with classical waves passing through the apparatus, an interference pattern will develop on the screen (I). If macroscopic billiard balls were passed through the set up, no interference (NI) is expected. But when a high intensity neutron beam is sent through; an interference pattern arises even though we view these as 'solid' particles. The concept of neutron interference is inconsistent with the classical concepts of particles, which don't interfere. But even when only one neutron is coming from the source at a time, an interference pattern still develops on the detection screen, which is predicted naturally from quantum mechanics but is unexplainable by classical mechanics (which would predict the NI pattern). This is because quantum mechanics is predicting the interference of potentials (each indistinguishable path is a potential) and not intensity or classical 'solid particle' propagation. But again in interferometry, all first order coherence is interference of single particles (including photons and neutrons) along indistinguishable paths (potentials). Paraphrasing Dirac, a photon can only interfere with itself. So two 'beams' do not actually interfere as in the classical description of OCT, which we will demonstrate quantitatively. Rather, although a more abstract concept, we can say the indistinguishable paths the particle has available can interfere (broadly, the path or field integral).

To illustrate the counter-intuitive relationship of indistinguishable paths and coherence, this Young's experiment will be examined with both environmental interactions (decoherence) and by moving the position of the detection screen (relative to the interactions). Later this principle of path indistinguishability (rather than combining beams) will be extrapolated to the interferogram of quantum OCT. In this description and figure 1, the E term will represent the environmental interactions/entanglements in the interferometer, such as collisions with a perpendicular electron beam. If we initially ignore the E terms (environmental interactions/entanglements), the pattern on the screen demonstrates interference that, as well will see, comes from the off-diagonal terms in the density operator. Now, if $E_1$ and $E_2$ are substantially different, such as when collisions with particles occur of significantly different momentum, the third and fourth terms disappear and the paths become distinguishable. The neutrons would then have different momentums depending on the path. Interference is lost in this simple example of environmentally induced decoherence (which occurs now that the paths are distinguishability based on momentum at the screen) [23-25]. The amount of difference between the E terms (environmental entanglement terms) affects the degree to which coherence (and interference) is lost (fringe visibility on the screen). If $E_1$ and $E_2$ are similar such as near identical particle collisions, the paths are still indistinguishable even though environment interactions occurred (paths indistinguishable at the screen/detection), and the interference pattern is maintained. Therefore, environmental entanglements do not necessarily lead to loss of interference if they are compensated for before measurement. Examples of this are the well-known quantum eraser and delayed choice experiments [26,27]. So path indistinguishability results in interference at the point of measurement and will still occur with environmental interactions as long as they occur in such a way that it cannot be determined which path the neutron took (a core but often unappreciated quantum mechanics principle).

So in this simple example examining path indistinguishability, again the interactions with E represents decoherence (which can be reversible or irreversible) while the interaction with the screen represents measurement (irreversible) as discussed in appendix A. Furthermore, the counter-intuitive and critical nature of indistinguishability can be illustrated if the screen is placed at A, B, or C with identical $E_1$ and $E_2$. If the screen is placed in either the A or C positions, an interference pattern will result but when in position B, interference is lost. This is because the paths are distinguishable at B as neutrons in each path have different momentum due to different environmental interactions. Even more poignant, if the screen is moved from B to C during the experiment, the interference pattern is recovered and decoherence reversed (paths go from distinguishable to indistinguishable). So interactions do not necessarily lead to loss of interference if they are compensated for before measurement is made. This recovered interference is also essentially the same phenomena behind the well-established quantum erasers and delayed choice experiments [26-27]. In addition, this reversibility of decoherence is used in the fields of quantum computers and information systems to preserve information, and may have a role in OCT [28]. The key aspect is that coherence is lost when the two paths are distinguishable <u>at measurement</u>.

*3E. Single Photon Interference Along Indistinguishable Paths, the Basis of OCT*

In a previous section, we represented the EM field (including the vacuum) in terms of quantum harmonic oscillators. We also qualitatively demonstrated single photon interference and path indistinguishability, including their role in first order correlations. In this section, we formally demonstrate why indistinguishable paths, particularly in linear interferometers such as those used with OCT and low coherence interferometry (LCI), are required for interference. We will be using single photons entering an interferometer one at a time for the initial analysis. So it will also be emphasized that most of the ranging achieved with OCT is predominately the linear superposition of large numbers of these single photon interferences (along indistinguishable paths). Again we are emphasizing that it is indistinguishable paths of a single photon and not 'recombining high intensity beams' that results in the interference with OCT. A single photon can only interfere with itself. In a subsequent section, through this analysis of single photon interference and the quantum harmonic oscillator, we will show a direct correspondence between the quantum and classical first order coherence function at high photon numbers. But before we make the comparisons with the correlation functions, we will examine single photon interference in the form of the classical OCT intensity interference equation above (1, 63-64).

In much of the text (and this section), monochromatic photons will be used to illustrative principles without loss of generality. Reviewing them again, the monochromatic single photon wavepacket (ignoring polarization) is define here:

$$|1_\omega\rangle \equiv \hat{a}^+(\omega)|0\rangle \qquad (24)$$

The 1 represents a single photon state. This has no classical analogy. Obviously, the notion of a localized monochromatic wave packet has no classical analogy. This is because classically, a monochromatic wave would be infinite in extent rather than a packet. It should be noted that an equivalent representation of the single photon, which also has no classical analogy, is the superposition of weighted Fock states. Since they are equivalent, the interested reader can find this representation elsewhere [equation 8].

But it will be demonstrated in subsequent sections that the ease of transition to the higher intensity sources of LCI and OCT (from single photon events) is straightforward. All linear first order coherence is a sum of single photon events (where you are simply summing the single photon events of different angular frequencies and amplitudes). This will be significantly more complex with second order coherence (ex: entanglement) where the bi-photon (dealt with in part II) can have very different behavior from a photon. Here a bi-photon is a photon pair (behaving as a single particle) that can only interfere with itself. While a first order coherent phenomenon almost always is the same for quantum and classical mechanics at high intensity, this is not true for second order correlations as will be discussed in section II.

This following section on single photon interference is partly based on extrapolating to OCT the pioneering work of Mandel for two sources [21]. From a quantum mechanical level, a beam splitter is not splitting intensities (as you can not split a single photon). So we need to discuss some of the quantum physics of a beam splitter. When discussing quantum noise later in the text, events at the beam splitter become important, much of which involves single photon interferences.

We limit ourselves to the simplest case of interference of two single-mode fields entering a beam splitter, as illustrated in figure 2, and being recombined after reflection in the arms of the OCT interferometer. The field is reduced such that a single photon enters one port of the beam splitter from the source and the other port vacuum fluctuations are entering (vacuum fluctuations from the source port can be ignored for reasons to be discussed). We will treat the vacuum fluctuations as homogeneous until the quantum noise section. The vacuum will therefore be labeled the 0 state out of convention but we already noted this represents nonzero energy.

In figure 2, the 'a' path is from the source and the 'b' path is the vacuum (later we will be using a second source to reduce vacuum fluctuation, a technique known as squeezing) [30,31]. The two potential photon paths after the beam splitter are 1 and 2. The quantum state after the beam splitter (again the entire state we start out with is a wave function with just a one photon state) is represented by:

$$|\psi\rangle = \alpha|1\rangle_1|0\rangle_2 + \beta|0\rangle_1|1\rangle_2 \qquad (|\alpha|^2 + |\beta|^2 = 1) \quad (25)$$

This is a single photon in a coherent superposition state between the two arms of the interferometer, with a probability of measurement $\alpha^2$ in arm one or in the other arm $\beta^2$. But the two possibilities are intrinsically indistinguishable. With respect to this analysis several points should be made:

1. We will deal with phase changes from the beam splitter in later sections. Reflection and transmission have different phases.
2. A topic that we will be addressing later in detail is the influence of vacuum fluctuations through the detector port (b in figure 2). But for analysis here, we will consider them insignificant. For first order correlations we will see these vacuum fluctuations primarily representing a noise source that operates through photon pressure at the ends of the interferometer (here the mirrors). However, they are not critical to the discussion of indistinguishable paths/coherence relationships at the beam splitter for first order correlations (though they will be for second order correlations).

To reiterate a point, with OCT first order coherence, photons are entering through one port of the interferometer from the source and vacuum fluctuations are entering through the other detector port. We want to find the density operator for the one photon system in equation 25 which is given by the general equation:

$$\hat{\rho} = |\Psi\rangle\langle\Psi| \tag{26}$$

The statistical information that describes the state of the quantized EM field is implicitly contained in its density operator. It should be stated that, unlike most operators, the density operator (ρ) may or may not have a carrot above it but it is always an operator. It is an operator in Hilbert space as described above. For indistinguishable paths, the density operator takes the expanded form:

$$\hat{\rho}_Q = |\alpha|^2 |1\rangle_1 |0\rangle_2 \langle 0|_1 \langle 1|_2 + |\beta|^2 |0\rangle_1 |1\rangle_2 \langle 1|_1 \langle 0|_2 + \alpha\beta^* |1\rangle_1 |0\rangle_2 \langle 1|_1 \langle 0|_2 + h.c. \tag{27}$$

where h.c. is the Hermititian conjugate. A <u>critical point is that even though there is only one photon, the last two terms are a superposition between paths that interfere (interference is occurring)</u>. Quantum mechanics predicts interference in this situation even with only one photon. The two occupation states are interference terms as they are not factorizable into independent components (3$^{rd}$ and 4$^{th}$ terms on the left of the equal sign). These interferences terms are the off diagonal elements of the matrix and are coherences. Classically their value would be zero but in quantum mechanics, the off diagonal terms have finite values.

On the other hand, when the paths are distinguishable the cross terms (α β* and β α*) go to zero. In principle, there exists, when the cross terms go to zero, an experimental set-up that allows determination of which arm the photon transversed. An example would be if the photon was frequency shift in one path or the other. Then the classical density operator has the diagonal form with off diagonal terms equal to zero (both alternatives are real and distinguishable):

$$\hat{\rho}_C = |\alpha|^2 |1\rangle_1 |0\rangle_2 \langle 0|_1 \langle 1|_2 + |\beta|^2 |0\rangle_1 |1\rangle_2 \langle 1|_1 \langle 0|_2 \tag{28}$$

This results in an incoherent classical mixture of states. In equation 28, off diagonal terms or coherence are lost because the last two coefficients go to zero so in principle, which path the photon took can be determined. In other words, the photon is in state one or two and not a superposition.

So these general results (with different constants) of equation 27 and 28 can be applied to either a Michelson's Interferometer (with OCT) or a Young's Interferometer. <u>So taking the Young's interferometer, equation 27 states that an interference pattern will occur on the screen, even though the photons are coming one at a time, when it can not be determined which path they came from. But if we can determine the path, equation 28, such as with a slight frequency shift in one arm, the interference pattern is lost.</u>

Now consider an arbitrary one-photon state within the given Hilbert space with the normalization terms expressed with density operators:

$$\hat{\rho} = \rho_{11}|1\rangle_1|0\rangle_2\langle 0|_1\langle 1| + \rho_{22}|0\rangle_1|1\rangle_2\langle 1|_1\langle 0| + (\rho_{12}|1\rangle_1|0\rangle_2\langle 1|_1\langle 0| + h.c.)$$

(29)

Again, the advantages of density operators over wavefunctions are that they can be extended to mixed states. Then we are going to divide equation 29 into probability of 27 or 28, the probabilities of quantum versus classical.

$$\hat{\rho} = P_Q\hat{\rho}_Q + P_C\hat{\rho}_C \qquad P_Q + P_C = 1$$

By equating matrix elements on both sides of this equation, we find that

$$\rho_{11} = |\alpha|^2$$
$$\rho_{22} = |\beta|^2$$
$$\rho_{12} = P_Q \alpha\beta^*$$

(30)

from which it follows that through 20, 21, and 22,

$$\alpha\beta^* = (\rho_{11}\rho_{22})^{1/2}\exp(i\arg\rho_{12})$$
$$P_Q = [\rho_{12}/(\rho_{11}\rho_{22})^{1/2}]\exp(-i\arg\rho_{12})$$
$$= |\rho_{12}|/(\rho_{11}\rho_{22})^{1/2}$$

(31)

It can be seen that indistinguishability is directly related to the cross density operators (coherence terms). When the $\rho_{12}$ (and it's complex conjugate) goes to zero, so does the indistinguishability (quantum) and vice versa. $P_Q$ is a measure of the degree to which the paths are intrinsically indistinguishable in the general quantum state $\rho$.

**4. OCT Correlation Functions (Classical and Quantum Mechanical)**

In the previous paragraphs we introduced the quantum field approach to the EM signal and single photon interferences in particular. In this section we will develop both classical and quantum OCT theory in terms of the coherence function. We will then compare both coherence functions derivations, which allows classical and quantum mechanical polychromatic comparisons between the approaches.

*4A. OCT Classical Correlation Functions*

OCT uses a Michelson interferometer where we will refer to the arms as reference (r) or sample (s) instead of 1 or 2. In the discussions that follow, it is assumed that all quantities are stationary. Stationary means that the time average is independent of the time of origin chosen (we are dealing with high intensities and not individual photons). The intensity of interference at the detector in a Michelson interferometer from a monochromatic source is given by (derived above, equation 1, and elsewhere)[1]:

$$I_D = 1/2\, I_{so} + 1/2\, I_{so}\cos\Theta = 1/2\, I_{so}(1+\cos\Theta) \qquad (32)$$

For the classical description we are describing interference in terms of intensity (or at times the electric fields), $I_{so}$ is the intensity of the source. When $\Theta$ (which can represent a time or distance mismatch) is equal to zero, the intensity at the detector ($I_D$) becomes $I_{so}$. When $\Theta$ is $\pm\pi$, the intensity at the detector is zero.

Next we examine equation 32 with a polychromatic source (without the distribution defined). So we are now going to modify equation 32 so that it contains an infinite number of wavelengths separated from each other by an infinitely small amount (1):

$$I_D = (1/2)\int_0^\infty I_{so}(k)(1 + \cos\Theta)dk \qquad (33)$$

Since $\omega$ and $k$ are proportional to one another, we can switch between them in the derivation as needed. Using a cosine identity and letting $\Theta$ equal to $kx$ where $x$ is the path length difference in the interferometer arms, equation 33 can be re-written as:

$$I_D = \frac{1}{2}\int_0^\infty I_{S0}(k)dk + \frac{1}{4}\int_0^\infty I_{S0}(k)(e^{ikx} + e^{-ikx})dk$$

$$= \frac{1}{2}I_{S0} + \frac{1}{4}\int_{-\infty}^\infty I_{S0}(k)e^{ikx}dk \qquad (34)$$

Here $I_{so}(k)$ is the sources power at a given value of $k$ and total power $I_{so}$ equals $\int_0^\infty I(k)_{so}$. Ignoring the first term after the equal sign, which is the DC signal, the autocorrelation function is given by:

$$G^{(1)}_{rs} = \int I_{so}(k)e^{ikx}dk \qquad (35)$$

The superscript 1 represents the first order coherence. This gets to the heart of how OCT and LCI function. OCT measures the classical autocorrelation function and uses it to represent backreflection. The backreflection data is then plotted in two dimensions in a manner analogous to ultrasound.

For OCT the source ideally is Gaussian that leads to a Gaussian autocorrelation function (optimal for plotting backreflection data). The classical autocorrelation function (in terms of the time delay) can also be represented more conveniently with respect to mismatch than the integral by:

$$G^{(1)}_{rs} = \langle E_r(t) E_s^*(t + \tau) \rangle_T \qquad (36)$$

Here $\tau$ is the time delay between the two interferometer arms, the Es are the complex random electric fields, and T is the time average. It can be normalized (so it is no longer a function of intensity) to the complex degree of coherence (classical) is:

$$g^{(1)}_{rs}(\tau) = G^{(1)}_{rs}/\sqrt{I_r I_s} \qquad (37)$$

Equation 34 can now be written more generally in terms of a correlation function:

$$I_D = (1/2)I_r + (1/2)I_s + \text{Re } g^{(1)}(\tau)_{rs}\sqrt{I_r I_s} \qquad (38)$$

This is a normalized function so that the real part of it has values from 0 to 1. The values of the degree of coherence is classified as follows:

$$|g^{(1)}_{rs}| = 1 \quad \text{coherent limit}$$
$$|g^{(1)}_{rs}| = 0 \quad \text{incoherent limit}$$
$$0 < |g^{(1)}_{rs}| < 1 \quad \text{partial coherence}$$

Before discussing the form of the coherence function for a Gaussian source power spectrum (the ideal OCT source), the concepts of a coherence time and length are addressed. These are critical to ranging with OCT and are dependent on the power spectrum of the source.

More detailed derivation of the classical coherence time can be found elsewhere [1]. Defining the width of a function is somewhat arbitrary (you have to define it), so we will use the power equivalent width:

$$t_c \equiv \pi^{-1/2} \int_{-\infty}^{\infty} |g(\tau)|^2 d\tau \qquad (39)$$

The coherence time can be described approximately as the time over which the EM field is relative constant and therefore can interfere with itself. The EM is relatively constant with respect to the amplitude and phase of the various frequencies. So, using classical language, if light down both arms of a Michelson interferometer travels the same transit time (within the coherence time), on recombination at the beam splitter interference occurs. However, if the relative delay (τ) between arms is greater than coherence time, no interference occurs. This allows ranging to be performed. The distance light travels during the coherence time is referred to as the coherence length ($l_c$) and is found by multiplying the coherence time by the speed of light. So if in an OCT system light from the reference and sample arm travel the same distance to within the coherence length, interference will occur. [We have seen from quantum mechanics, the more precise definition is that if the paths available to a photon are indistinguishable, including path length, single photon interference occurs]. The intensity of interference is used to represent backreflection intensity and the two-dimension backreflection profile is used to give structural detail (analogous to ultrasound).

For monochromatic light, the coherence length is infinite. As described elsewhere, when the source spectrum is Gaussian, optimal ranging is achieved. When the source spectrum (S) has the form of a Gaussian function, it can be written as:

$$S(\omega - \omega_0) = A(\frac{2\pi}{\sigma_\omega^2})^{1/2} \exp[-\frac{(\omega - \omega_0)^2}{2\sigma_\omega^2}] \qquad (40)$$

From 35 and 37, we can obtain the normalized correlation function for a Gaussian source spectrum (derived in detail elsewhere):

$$g(\tau) = \exp(-\sigma^2 \tau^2 / 2) \exp(i\omega_0 \tau) \qquad (41)$$

Or in terms of the coherence time ($\tau_C$):

$$g^{(1)}(\tau) = e^{-i\omega_0 \tau - \frac{\pi}{2}(\tau/\tau_c)^2} \qquad (42)$$

We will see that quantum mechanics generates the same results for the coherence function. But unlike the classical explanation of interfering intensities, quantum mechanics also effectively explains phenomena like single photon interference. But, in part II, we will see where the behavior of second order correlations at large photon numbers can deviate from classical predictions. These second order phenomena offer the potential to expand the diagnostic potential of OCT.

*4B. OCT Quantum Correlation Functions*

In the previous section, we demonstrated the classical correlation function used with OCT ranging. This classical function does not account for quantum effects such as single photon interference (which is actually the basis of all first order correlations), demonstrating that it is a limited description. In this section, we will develop a correlation function that is quantum mechanical in origin, yet still incorporates the features of classical mechanics. The quantum correlation function can be expressed with the use of electric field operators, Fock states, and density operators (rather than the intensity form of the previous sections). It will use the quantum harmonic oscillator model. Though we will begin with single photons we will progress to large photon numbers in a Gaussian frequency distribution.

We will start with the electric field operators. $\hat{E}^+(r)$ can be viewed as analogous to the analytical electric field in classical mechanics. Incorporating all constants into a constant K, the $\hat{E}^+(r)$ at a position r in either arm of the beam splitter can be expressed as:

$$\hat{E}^{(+)}(r_j) = K\hat{a}_j \quad j = 1,2$$

(43)

This is an extension of equation 21. The value of K is defined below but the focus is (of equation 43) the linear relationship between the operators. The operator $\hat{E}^-(r_j)$, the complex conjugate of $\hat{E}^+(r)$, is proportion to the creation operator $\hat{a}^+_j$ (photon emission) so that the total electric field operator is given by:

$$\hat{E}(r,t) = \hat{E}^{(+)}(r,t) + \hat{E}^{(-)}(r,t)$$

(44)

[In a more complete description below, the electric field operator works on the density operator to yield the ensemble average of the field.] Photon correlations are measured during detection, where a photon is lost from the field. In other words we are interested in the OCT field at the detector. So the derivation is primarily in terms of the $\hat{E}^+$ operator (and therefore proportional to the annihilation operator) on the field as photons are being destroyed at the detector. The initial state before absorption is i (not to be confused with the imaginary number i) while the final state is f. We will replace these with wave functions (for pure states) and density operators (pure states and mixtures) shortly. Therefore, the transition rate of an absorbing detector atom (an intensity) is given as per Glauber by [16,17,32]:

$$u_1(r,t) = \left| \langle f | E^{(+)}(r,t) | i \rangle \right|^2$$

(45)

The quantity is squared because u(r,t) is a photon counting rate (absorption rate for a given final state) or the expectation value of the operator. It is obviously proportional to the intensity measured at large photon numbers. So again, we are deriving a correlation functions in quantum mechanical formalism using field operators to yield the detected fields. Since the distribution of final states in equation 45 can almost never be measured, we can express the detection rate in terms of the field operators and initial states only:

$$u_1(r,t) = \sum_f |\langle f|E^{(+)}(r,t)|i\rangle|^2$$

$$= \sum_f \langle i|E^{(-)}(r,t)|f\rangle\langle f|E^{(+)}(r,t)|i\rangle$$

$$= \langle i|E^{(-)}(r,t)E^{(+)}(r,t)|i\rangle \tag{46}$$

The summation is for all possible f for each given i. The utility of the equation can be further extended to include the probability of all possible initial states such that:

$$u_1(r,t) = \sum_i P_I \langle i|E^{(-)}(r,t)E^{(+)}(r,t)|i\rangle \tag{47}$$

Using the initial (i) and final states (f) are convenient for expressing the theory but now we want to express states in terms of the density operator. We will use the convention of a wavefuntion to represent a pure state and the density operator is the sum of the pure states (where the sum can be just one pure state so a pure density operator). Then the expectation value can now be expressed in terms of the density operator (which encompasses all initial states and their coherence) using the general derivation:

$$\langle \hat{O} \rangle = \sum_n p_n \langle \Psi_n|\hat{O}|\Psi_n\rangle$$

$$= \sum_{ij}\sum_n \langle \Psi_n|i\rangle\langle i|\hat{O}|j\rangle\langle j|\Psi_n\rangle \tag{48}$$

$$= \sum_{ij} \hat{O}_{ij}\hat{\rho}_{ji} = Tr[\hat{O}\hat{\rho}]$$

Here $\hat{O}$ is an operator, Tr is the trace, $\rho$ is the density operator, the wave function is a pure state, and $p_n$ is a probability function. The statistical information that describes the state of the quantized electromagnetic field is implicitly contained in its density operator. So now combining equations 47 and 48 we obtain for the expectation value:

$$u_1(r,t) = Tr[\rho E^{(-)}(r,t)E^{(+)}(r,t)] \tag{49}$$

So we have a quantum way of assessing the field at the detector. With OCT we are interested in <u>comparing</u> the analytic field with itself in terms of either a distance or time delay. In practice, what is measured through equation 49 is a statistical average, the correlation function. So the first order correlation function (the autocorrelation function) can be defined as (ex: in terms of time delays):

$$G^{(1)}(\mathbf{r}_1, \mathbf{r}_2; t) = Tr[\rho E^{(-)}(\mathbf{r}_1, t_1) E^{(+)}(\mathbf{r}_2, t_2)]$$
$$= \langle E^{(-)}(\mathbf{r}_1, t_1) E^{(+)}(\mathbf{r}_2, t_2) \rangle \quad (50)$$
$$= \langle E^{(-)}(\mathbf{r}, t) E^{(+)}(\mathbf{r}, t+\tau) \rangle$$

where $\tau$ is the difference in time propagating at the detector after traveling through the interferometer. The complex degree of coherence or the normalized first order degree of coherence (with respect to delay) is (normalized for the intensity):

(51)
$$g^{(1)}(t) = \frac{\langle E^{(-)}(t) E^{(+)}(t+\tau) \rangle}{\sqrt{E^{(-)}(t) E^{(+)}(t) E^{(-)}(t+\tau) E^{(+)}(t+\tau)}}$$

This can be further simplified by dividing out K and expressing it in terms of the annihilation operators (here for a radiation field in a single mode):

$$G^{(1)}(\tau) = \langle a^+(t) a(t+\tau) \rangle \quad (52)$$

$$g^{(1)}(\tau) = \frac{\langle a^+(t) a(t+\tau) \rangle}{\langle a^+ a \rangle}$$

We have now developed the OCT coherence function from both a classical and quantum mechanical basis. This gives us insight to the limitations of the classical OCT coherence function discussed in the next section.

*4C. Relationship Between the Quantum and Classical OCT Correlation Functions*

We have presented the classical and quantum mechanics of linear interferometry with first order coherence, the latter using the quantum harmonic oscillator and single photon interference. In the limit of high photon counts, unlike second order correlations, they demonstrate the same behavior so this is of little interest (we are treating position probability amplitude of the target as a separate issue). But at low photon counts the quantum behavior of first order correlations becomes very significant. As a building block for both the remainder of the paper and subsequent work, we will demonstrate that the classical OCT intensity interference equation (35), autocorrelation function, and Gaussian autocorrelation function can be reproduced by the accumulation of single photon interferences alone (representing the photon field as quantum harmonic oscillators).

In the limit of high photon numbers, almost all linear quantum first order correlations reduce to classical first order correlation results. This is not true for non-linear interferometry, as ours and other groups have shown [61,63]. Again, it is also not true for second order correlations as will be seen in the part II that will likely be an important area for advancing OCT's diagnostic capabilities.

We will go back to a single photon pure state to generate the coherence function. It represents a superposition of the single monochromatic photon in either arm of the interferometer. Again, we are ignoring phase and polarization changes by the components as they do not influence the theory:

$$|i\rangle = \frac{1}{\sqrt{2}}(|1\rangle_1|0\rangle_2 + |0\rangle_1|1\rangle_2) \tag{53}$$

From the previous section, the first order correlation function, for a Michelson interferometer with a given phase delay, is given by:

$$G^{(1)}(r,r') = \langle i|E^{(-)}(r',t)E^{(+)}(r,t)|i\rangle \tag{54}$$

Assuming an absorptive process (the final state of the field is a monochromatic vacuum state), combining 53 and 54:

$$\begin{aligned}G^{(1)}(r,r') &= \langle i|E^{(-)}(r',t)|0\rangle\langle 0|E^{(+)}(r,t)|i\rangle \\ &= \Psi_\varepsilon^+(r',t)\Psi_\varepsilon(r,t)\end{aligned} \tag{55}$$

The electric field operators are given by [derived in reference 16]:

$$\hat{E}^{(+)}(r,t) = (\hat{E}^{(-)}(r,t))^\dagger$$

$$= i\int_0^{+\infty} (\frac{\hbar\omega_k}{2\varepsilon_0 v})^{\frac{1}{2}} \bar{e}_k \hat{a}_k \exp[i(nk \cdot r - \omega_k t)]dk \tag{56}$$

This allows the following wavefunction to be calculated (from equations 55 and 56):

$$\begin{aligned}\Psi_\varepsilon(r,t) &= \langle 0|E^{(+)}(r,t)|i\rangle \\ &= C(e^{ikr} + e^{ikr'})\end{aligned} \tag{57}$$

The two exponentials result from propagation in the two arms of the interferometer as in equation 56 and their form (plane wave, spherical wave, etc.) is absorbed in the constant C. In addition, all constant terms not critical to the discussion are placed in the constant C. Now using equation 56 and 57, through a common trigonometry identity, the exponential can be converted into a cosine such that:

$$\begin{aligned}G^{(1)} &= D(1+\cos[k(r-r')]) \\ &= D(1+\cos[k\Delta l]) \\ &= D(1+\cos\theta)\end{aligned} \tag{58}$$

The D term simply represents the C constant times numerical constants (not needed for the discussion) added in the conversion. The term $\Theta$ has been used here to represent the phase mismatch (time or distance) at the point of detection between arms. This mismatch can be in either the time or spatial domain. Equation 58 yields the classical interference equation (derived above) for the OCT Michelson interferometer (assuming a 50:50 beam splitter, perfect reflection in both arms, and monochromatic light) where $I_{SO}$ is the intensity of the source [1,30]:

$$\begin{aligned}I_B &= \frac{1}{2}I_{S0} + \frac{1}{2}I_{S0}\cos\theta \\ &= \frac{1}{2}I_{S0}(1+\cos\theta) \\ G^{(1)} &= B(1+\cos\theta)\end{aligned} \tag{59}$$

So both the quantum and classical OCT coherence function express first order coherence as a constant times 1 plus the cosine of the mismatch angle. The quantum coherence function was generated from single photon interference rather than the 'combination of two beams'. So whether photons are coming one at a time or at a high intensity beam, equations 33 and 59 give the same intensity variations with mismatch. This emphasizes that all first order coherence (with a linear interferometer) are the linear superposition of single photon interferences. The same holds true for polychromatic high intensity ranging with OCT as we will see in the next few paragraphs, but we will express the comparison in terms of the correlation function.

We now move forward in a relatively straightforward manner to a Gaussian autocorrelation function where single photon interferences use photons with a Gaussian frequency distribution. Part of this relates to the interferometry work of Saleh and Teich, though that did not involve OCT or LCI [56]. Three points, important to this derivation, have already been made. First, based on the discussion to this point we are dealing with broadband high intensity interferometry that is a linear sum of the superposition of single photon interference events. Second, interference can be described in terms of density and electric field operators, using the Fock state (annihilation operator) basis. Third, with both the quantum mechanical as well as the classical approach to OCT, the autocorrelation function is proportional to the Fourier transform of the source spectrum. So a Gaussian source spectrum leads to the optimal autocorrelation function (Gaussian).

The OCT signal can be constructed from wave-packet modes that are built from weighed superpositions of the monochromatic modes of the field. The analysis will be in one dimension. A polychromatic annihilation operator can be constructed from the various annihilation operators and their frequency distribution $\{\varepsilon(\omega)\}$:

$$\hat{A}^+(\varepsilon) = \int_0^\infty \varepsilon(\omega)\hat{a}^+(\omega)d\omega \qquad (60)$$

As above, this is proportion to the electric field operator as discussed earlier. In OCT the ideal normalized distribution function $\varepsilon(\omega)$ is Gaussian based on the source characteristics and is given by:

$$\varepsilon(\omega) = (2\pi\sigma^2)^{-1/4} \exp\left[-\left[\frac{\omega-\omega_0}{2\sigma}\right]^2\right]\exp(-i\omega t_0) \qquad (61)$$

We define $\sigma$ as the full width half maximum of the spectrum. The distribution function can be normalized via:

$$\int_0^\infty |\varepsilon(\omega)|^2 d\omega = 1 \qquad (62)$$

We now reintroduce the state initially with the monochromatic single photon wavepacket (as above):

$$|1_\omega\rangle \equiv \hat{a}^+(\omega)|0\rangle \qquad (63)$$

This has no classical analogy but plays an important role in quantum mechanics. Using the definition of the wavepacket operator, the OCT polychromatic single photon wavepacket is given by:

$$|\Psi_\omega\rangle \equiv \hat{A}^+(\varepsilon)|0\rangle = \int_0^\infty \varepsilon(\omega)|1_\omega\rangle d\omega \qquad (64)$$

So again we are dealing with single photons. Again, the autocorrelation function is given by:

$$G^{(1)}(t,\tau) \equiv Tr\left[\hat{\rho}\hat{E}^{(-)}(t)\hat{E}^{(+)}(t+\tau))\right] \qquad (65)$$

Where the density operator is again given by (using the Gaussian distribution of equation 61):

$$\hat{\rho} = |\Psi_\omega\rangle\langle\Psi_\omega| \qquad (66)$$

The normalized coherence function is given by:

$$g(\tau) \equiv \frac{\int_{-\infty}^\infty G^{(1)}(t,\tau)dt}{\int_{-\infty}^\infty G^{(1)}(t,0)dt} \qquad (67)$$

The solution for this complex degree of coherence (for the Gaussian distribution) is therefore given by:

$$g(\tau) = \exp(-\sigma^2\tau^2/2)\exp(i\omega_0\tau) \qquad (68)$$

This is the same result as for the classical derivation. But it was done using a density operator of single photon interferences with a Gaussian frequency distribution and a quantum harmonic oscillator representation (equation 11). The coherence time again is defined as:

$$t_c \equiv \pi^{-1/2}\int_{-\infty}^\infty |g(\tau)|^2 d\tau \qquad (69)$$

For a Gaussian function then, the coherence time can be rewritten as:

$$t_c = \frac{1}{\sigma} \qquad (70)$$

So for OCT (ignoring constants for phase and polarization changes from optical components), ranging was effectively modeled using a quantum field approach was performed using a quantized harmonic field and single photon interference. We will see how this yields different results for noise in the system than a classical approach.

**5. Application of Quantum Optics to Advancing OCT: Quantum Noise Reduction**

In the previous sections, we built the foundations for advancing OCT through quantum optics. In the remainder we will focus on the practical application of this to OCT imaging, primarily where single photon interference provides insights into quantum noise reduction. An additional reason for writing this section is, in addition providing a first description of quantum noise sources in OCT, it gives a sense of the degree of experimental and theoretical work that still needs to be done specific to OCT.

*5A. General*

The primary reason for our group studying quantum OCT is second order correlations and spread of the position probability density, which are dealt with in part II. We believe these areas will expand OCT's diagnostic imaging capabilities. But there are other areas where quantum OCT can lead to advances in diagnostic capabilities. In this paper, we will provide an example where first order correlations give insights to improve diagnostic imaging and where research effort is needed. This is the area of quantum noise sources and the quantum noise limit. This determines the theoretical limit of the OCT signal to noise ratio (SNR) and it is primarily a function of single photon interferences/first order correlations. In the field of gravitational wave research (which will provide insights into OCT), field strengths with Michelson interferometer measurements are very low so much effort has successfully gone into reducing noise below the standard quantum limit (SQL). We will discuss quantum noise first qualitatively and the quantitatively. But work in both theory and application is needed in OCT quantum noise beyond that extrapolated from the gravitational field work, which will be discussed.

People often use the phrases 'optical shot noise' and 'optical quantum noise' synonymously. But the definition of optical shot noise, whose origins dates back to the 50's, the author has found to be inconsistent in the literature so it will not be used here. As our group has done with previous work, for OCT the term optical quantum noise will be used instead. There are also many misconceptions about optical quantum noise. It is commonly stated that it is the "the process of random absorption of the EM field by the quantized detector atoms" (which is how some define shot noise). This ignores the fact the EM field is already quantized so quanta are being absorbed by the detector and not a continuous field. For OCT, we will define quantum noise as either optical energy or target position fluctuations that decrease the accuracy of ranging information and are not predicted by classical mechanics. Remember from the discussion above that photons entering the beam splitter individually (we are using a Michelson interferometer), even at irregular intervals, will result in single photon interference when paths are indistinguishable. In other words, first order coherence is not dependent on constant photon number per unit time. This can be seen from equations 55,57-58. Quantum noise is generated primarily from photons that travel distinguishable paths (we will be focused here on single and not bi-photons as above). This can be viewed as photons that either do not undergo interference (but are detected) or are applying asymmetrical photon pressure to each arm (increasing position probability amplitude uncertainty in that arm and therefore variable ranging [$\Delta z$]). These turn out to be equivalent. The position probability uncertainty in the target is a complex issue. The position probability uncertainty of the target, using a mirror as an example, is complicated by the fact the object can behave as if it is one macroscopic object or much smaller subdomains (even individual molecules). The latter has a much greater amplitude than the former. Factors influencing the position probability amplitude include effective mass, decoherence, second order correlations, and entanglement. So this will be dealt with briefly below but predominately in part II.

In this section, we shall be concerned primarily with two types of quantum errors. These are photon counting errors (PCE) and vacuum fluctuations entering through the detector port (in addition to some discussion of position uncertainty). They are explained in detail below. These two are believed to be the two major mechanisms of quantum noise in the EM field.

Before looking at specific quantum noise sources, we can just examine the theoretical limit posed by Heisenberg's uncertainty principle. The standard quantum noise limit (SQL) is the limit for the conjugate pair (ex: intensity-phase, momentum-position) by the uncertainty principle. But it should be remembered that this can be reduced in one of a given quatrature pair at the cost of increased uncertainty of the other, called squeezing [36-38]. The SQL is approximately equal to the PCE and radiation-pressure from vacuum fluctuations entering the detector port of the beam splitter [36-38]. A general approximation is that, for a measurement of duration τ, the probable error in the interferometer's determination of Δz (mirror position) can be no smaller then the SQL, which is approximately [36-37]:

$$(\Delta z)_{SQL} = (2\hbar\tau/m)^{1/2} \qquad (71)$$

This limit results from a direct extension of Heisenberg's uncertainty principle applied to the quantum mechanical evolution of a free mass. Three initial points with regard to this limit. First, this is intended to represent the minimal approximate quantum noise. The quantum noise can be significantly higher. As one example, backreflection off an inefficient detector surface where light energy is mixing with the vacuum fluctuations at the beam splitter can increase noise above the predicted SQL. Second, again for a given quatrature pair, one of the pair can be reduced below the SQL at the expense of the other. Third, there are situations where the mass (m) can't (or shouldn't) be represented by the entire target mass (ex: mirror) but individual molecules or subdomains. This is dealt with in part in the next section and more detail in part II.

*5B. Quantum Noise Source: Qualitative*

a. Position Probability Amplitude Uncertainty

Before discussing vacuum fluctuations and PCE as quantum noise sources, we will briefly touch on position probability amplitude uncertainty. When thinking about an object's position for ranging, for example mirrors in both arms, we tend to view them as classical with well-defined positions and borders. But each molecule and subdomain, as well as the entire object, have a position probability amplitude and are not a well-defined position.

We normally consider position uncertainty of an entire entity by the uncertainty principle, using it either in the form of the standard quantum limit (above incorporating detection time) or the instantaneous traditional Heisenberg representation dx ≥ h / 4πmdv where dx is the position uncertainty, m is the mass, and dv is the velocity uncertainty. So the tendency is to think of an object like a large mirror to have an almost unappreciable uncertainty. However, it is far more complicated than this which is rarely considered. First, many large objects (like mirrors) often behave as consisting of subdomains or even a collection of individual molecules (many rather than a single component). How the object behaves can be influenced by the field (ex: second order correlations can cause a mirror to behave as small domains) [3]. When the object does not behave as a single unit, the mass and uncertainty of these regions are much larger than those anticipated for the whole mirror. Second, as demonstrated by a seminal study of decoherence, Joos and Zen showed while decoherence localizes (establishing one eigenvalue) it does so in individual units or subdomains [4]. Therefore, while an individual particle or entity can be localized by decoherence if they are part of a larger object like a mirror, the larger object (summed over these smaller subsystems) can be spread over a wider region of space (much higher uncertainty). Third, in studies from the group at BU using entangled photons and thermal second order correlations by our group, interference was noted between reflectors separated by millimeters even though the coherence lengths were less than 20 µm due to the spread of position probability density [3,5]. In both cases, it was felt a very broad spread of the position probability amplitude was taking place due to second order correlations [5,13,26-27][3,5,18, 43]. Because second order correlations are more prevalent with OCT than gravitational wave measurements because the former uses a broadband thermal source, they will have a greater impact in OCT. This includes spreading the position probability density. Again, this source of quantum uncertainty (ranging errors) will be discussed in more detail in part II as it is more complex than noise sources described below. An important point addressed will be, in scattering theory, with position probability amplitudes being so large, how can they be ignored.

b. Vacuum Fluctuations

Vacuum fluctuations were introduced earlier as the polychromatic lowest energy limit of the field. Most of the initial quantum noise analysis will be of a monochromatic vacuum. Qualitatively, in OCT, vacuum fluctuations cause noise primarily by asymmetric photon pressure on the end mirrors (targets) in each arm, leading to position uncertainty (shown quantitatively below). Vacuum fluctuations have been shown to be a major noise/error source in a Michelson interferometer, with substantial contributions coming from work in gravitational wave research. But this work generally uses a coherent monochromatic source that has differences when compared to OCT. With OCT, the use instead of a broad bandwidth Gaussian source needs to be taken into account.

With a previous body of work demonstrating the significance of noise (in interferometers) caused by vacuum fluctuations, it is somewhat surprising resistance exists in the OCT field accepting their relevance. It is even further surprising because vacuum fluctuations are of importance in other areas of optics, for example; being the source of spontaneous light emission, photons from spontaneous parametric down converted (SPDC) sources, and the Casmir effect [43-46]. They are even postulated to be the source of dark energy accelerating the entire universe, yet the general feelings is that they are too microscopic to be relevant to OCT imaging [47-49]. Over three decades of evidence would argue otherwise.

As already described the vacuum consists of non-zero energy modes (equation 12 and 13), a direct result of the uncertainty principle and modeled in this paper by quantum harmonic oscillators. The energy and characteristics of the fluctuations depend on the modes that are present in the vacuum at any space-time point, as well as their interaction with non-vacuum modes [45, 48-49]. The frequencies of the vacuum energy is in general a function of the size and shape of the volume they are contained in (allowable modes), with the Casmir effect being a prime example, where not all vacuum frequencies are around between metal plates [45-47]. With OCT or any Michelson interferometer, vacuum fluctuations have their greatest impact from the energy fluctuations enter the detector port of the interferometer beam splitter. These leads to quantum noise when combined with out of phase source light (described quantitatively below). This is unlike vacuum fluctuations entering through the source port that have minimal effect.

With respect to entering the detector (exit) port, the significance lies in the interaction with the source light entering through the source port and vacuum fluctuations, where the phase difference between reflection and transmission in the beam splitter leads to noise. If the fluctuation has the right phase to increase the intensity from the source in one arm, it decreases the intensity in the other arm creating distinguishable paths, which results in noise (described quantitatively below) [50-52]. Asymmetrical radiation pressure then leads to path length uncertainty (Δz), the force of which is proportional to the square root of N [39,40].

c. Photon Counting Error (PCE)

Again, the other major type of quantum noise error is associated with the fluctuations in the photon count or the time average of the light (PCE) [33, 42]. In spite of this the origin of PCE still is somewhat controversial and in many papers on quantum noise, the origin is not addressed but the existence just taken for granted. It is <u>not</u> the random emission of photons from the source or a randomness of the detection process, as is often thought. In general, laser sources approximate a Poisson distribution while thermal sources used in OCT follow Bose-Einstein photon statistics. Although this guarantees fluctuations in photon number per unit time, we already demonstrated first order correlations are the summation of single photon interferences along indistinguishable paths [33,42]. Therefore, they only cause fluctuations in the autocorrelation function when the pathlength mismatch is greater than the coherence length, leading to distinguishability (photons coming from one arm or the other). This is not the major source of PCE when mirrors are targets in each arm. However, when dealing with tissue, which has scattering from different depths, backreflection from areas of mismatch make significant contributions to the PCE, which has not been studied in detail and is not as significant in the gravitational wave research area (mirrors both arms).

But also a substantial contributor of PCE is optical components in each arm of the OCT interferometer making paths indistinguishable. If photons came to an ideal detector perfectly spaced in time, the distribution would be perfectly sub-Poisson with no bunching or anti-bunching (in the quantitative description below, we will lay the noise on this hypothetical distribution). Even with this hypothetical source, passing through a beam splitter (other than the initial) or reflecting off the mirror/fiber bends (or a variety of other components) leads to a Poisson-like distribution (so some of the photons will be bunched out of random chance). These changes are asymmetrical in each arm so single photon interference is not produced and the system behaves as if each photon is traveling down one arm or the other (no superposition occurs). In other words, each arm is distinguishable so photon pressure is unequal and therefore Δz changes. So components leading to asymmetrical photon pressure down either arm are a major source of PCE. Irrespective of the predominate source of PCE in the OCT interferometer, PCE is universally recognized as a source of quantum noise even if the exact origin is not completely understood or agreed upon.

*5C. Quantitative Quantum Noise*
a. General

This section provides a quantitative evaluation of OCT quantum noise through two approaches, one at the beam splitter and one at the detector for reasons discussed. We will use as a foundation work from the gravitational wave field where Caves focused on analyzing noise from within the interferometer while Loudon (later by Ben-Aryeh) looked at combined quantum noise at the detector. Since the effect of terrestrial gravitational waves on the interferometer arm length is extremely small, it is very difficult to achieve gravitational wave detection and much effort is spent in overcoming all the technological problems related to extremely accurate Michelson interferometer measurements (we will discuss aspects of OCT distinct from the gravitational work). The Caves approach both demonstrates how vacuum fluctuations lead to noise (at the level of the beam splitter) and further demonstrates how 'squeezing' works in reducing these ranging errors. Squeezed vacuum Michelson interferometers (often using a second source in the detector port) are commonly used in the study of gravitational waves.

But when looking to measure total quantum noise, it is more ideal to examine it at the detector (where all quantum noise contributions are added). This is the basis of the approaches by Loudon and Ben-Aryeh. Contributions from PCE and vacuum fluctuations are not separated (and they argue can not accurately be separated) and the combined noise is modeled to be measured at the detector. One of the critical parts of the approach is the noise is placed on an ideal sub-Poissonian (evenly spaced) distribution by artificially introducing a non-linear Kerr effect.

The Caves, Loudon, and Ben-Aryeh approaches were developed for studying gravitational waves. But several significant differences exist between a gravitational wave interferometer and an OCT system. These include the large differences between mirror masses (gravitational interferometers have very large mirrors), light used (gravitational studies use monochromatic sources with high intensities), and OCT generally uses a multi-layer object in the sample arm. The much smaller mirrors of OCT, for example, result in higher position probability uncertainties. This is compounded by the fact it is not monochromatic as the light (near infrared) is polychromatic thermal light (Gaussian distribution) resulting in more complex photon pressure effects. Finally, the fact the target in the sample arm has reflections outside the coherence length leads to PCN.

b. Vacuum Fluctuations at the Beam Splitter Causing Noise

In this description at the beam splitter, we begin with light coming into the beam splitter from the source port and vacuum fluctuations entering from the detector port. Technically, it is a common statement that photons enter the beam splitter. No measurement is taking place so strictly speaking the terminology should be with respect to the potential and not actual photons (the photon are only being measured at the detector). We will continue with this common approach recognizing no measurement has actually taken place till the detector.

Returning to the model, in one sense the vacuum fluctuations can be viewed as half photon energy but unlike a photon, the vacuum can add energy to a source photon (an interaction occurs). The beam splitter is represented as in Caves, which uses the following transformation for the beam splitter [36,37]:

$$\hat{b}_3 = 2^{-1/2} e^{i\Delta} (\hat{a}_1 + e^{i\mu}\hat{a}_2)$$
$$\hat{b}_4 = 2^{-1/2} e^{i\Delta} (\hat{a}_2 - e^{-i\mu}\hat{a}_1)$$

(72)

Again we are combining source photons with vacuum fluctuations through the two ports. Equations demonstrate the different phase shifts from reflection and transmission.

So each input has different phases for each output arm. The two output arms are no longer in phase so paths are distinguishable. Here $\hat{a}_1$, $\hat{a}_2$ are the annihilation operators of the beamsplitter's two 'in' modes (source and vacuum) and $b_3$, $b_4$ are the operators for the two 'out' modes (two arms of the interferometer), while µ and Δ are the relative and the overall phase-shifts, respectively. The overall (global) phase shift will be ignored here since (as is often the case in quantum mechanics) it does not affect our results. Note the exponential with the µ term has different signs for each output port, resulting in the phase difference from 'mixing' of the vacuum and source EM, with distinguishability of the b operators.

The state leaving the beam splitter can also be represented by:

$$\hat{b}_3 = \frac{1}{\sqrt{2}}(\hat{a}_1 + i\hat{a}_2) \qquad \hat{b}_4 = \frac{1}{\sqrt{2}}(i\hat{a}_1 + \hat{a}_2)$$

(73)

Now from Caves, the difference in momentum transfer between each arm (which leads to distance uncertainty) can be given by the operator:

$$\hat{p} \equiv (2B\hbar\omega/c)\left(\hat{b}_4^+\hat{b}_4 - \hat{b}_3^+\hat{b}_3\right)$$
$$= -(2B\hbar\omega/c)\left(e^{i\mu}\hat{a}_1^+\hat{a}_2 + e^{-i\mu}\hat{a}_2^+\hat{a}_1\right)$$

(74)

Here we have expressed it explicitly in terms of the input $\hat{a}$. The B is the bounces off the reflector. From the second term, we see that this difference between arms occurs with differences in the phase from the beam splitter due to transmission versus reflection. So in the Caves analysis, the phase differences in energy between the source and vacuum ports (final term in parenthesis in equation 74) leads to path distinguishability (and therefore momentum differences). The different photon pressure alters both the position probability density of the target and the phase of returning photons. This creates optical quantum noise (within otherwise the coherence length) as opposed to vacuum fluctuations entering through the source port (which are in phase between arms). Again, in this analysis first order coherence, as discussed earlier in the paper, is the linear summation of the single photon interferences.

Squeezing the vacuum at the detector port will not be dealt with here in detail. However, it can be imaged that by altering the vacuum fluctuation, such as with a second source in that port, error can be reduced dramatically for either amplitude or phase measurements.

c. Combined Quantum Noise at the Detector

The previous analysis provides insights into ranging deterioration from vacuum fluctuations at a beam splitter and provides ways to reduce their influence as a noise source (ex: squeeze states). This is important from a mechanistic analysis, and also provides a way to reduce noise. However, from a practical standpoint when measuring total quantum noise we are interested in total amplitude quantum noise at the detector that leads to ranging errors. In addition, there is criticism that the vacuum fluctuations and PCE can't be completely separated from each other. We will adapt an approach from Loudon (later modified by Ben-Aryeh) to OCT. In the previous section we began with the EM field before the beam splitter while here, we ignore that interaction and focus on events after the 'beam is initially split' (more accurately on the sum of the two single photon paths). Here, we will use the one mode boson operators $\hat{b}_1$ and $\hat{b}_2$ as exiting the beam splitter through both interferometer arms (i.e. we start after the first pass through the beam splitter).

The states exiting the beam splitter on the return trip (headed back to the detector and source) are then given by the operators $\hat{d}$ and $\hat{e}$. Photon pressure interacting with the mirror in each arm, both from vacuum fluctuations and PCE, is represented as a non-linear Kerr effect for reasons described in detail by Haus [53,54]. But basically we are using the Kerr effect to create a sub-Poissonian (evenly spaced) distribution on which to overlap the quantum noise sources. The analysis takes into account the respective path lengths $Z_1$ and $Z_2$ (reflection off the distal mirrors), the sub-Poissonian field, and the quantum noise sources:

(58)
$$\hat{d} = \frac{1}{\sqrt{2}}[\exp(ikZ_1 + iC_1\hat{b}_1^+\hat{b}_1)\hat{b}_1 + \exp(ikZ_2 + iC_2\hat{b}_2^+\hat{b}_2)\hat{b}_2]$$

$$\hat{e} = \frac{1}{\sqrt{2}}[\exp(ikZ_1 + iC_1\hat{b}_1^+\hat{b}_1)\hat{b}_1 - \exp(ikZ_2 + iC_2\hat{b}_2^+\hat{b}_2)\hat{b}_2]$$

where k = 2π λ is the wavenumber of the one-mode EM field. The parameters $C_1$ and $C_2$ depend on the properties of the targets in each arm (example free floating versus harmonic mirror), and their explicit evaluation is not discussed here. The transformations include the exponentials with number operators that are higher powers, which illustrates the nonlinearity of the Kerr-type interaction used to form the quantum noise free basis. We simplify for now the present treatment by assuming C1 = C2 as is done in the case of gravitational interferometers. However, with a small mirror such as those used in OCT (as well as the multi-layer target in the sample arm) this approximation may not hold (and needs to be evaluated experimentally), in which case photon pressure induced noise would be dramatically increased (very large differential phase shifts). This is one area where work needs to be done with OCT to both understand and control OCT specific quantum noise.

First order or single photon interference depends on the length difference $Z_2 - Z_1$, but one may add to this parameter any additional interferometer effects at constant phase difference including birefringence and inefficient reflection/transmission. The photon number in one of the output ports of the interferometer is then given by:

$$\hat{d}^+\hat{d} = \frac{1}{2}(\hat{b}_1^+\hat{b}_1 + \hat{b}_2^+\hat{b}_2 + \{\hat{b}_1^+\exp[ik(Z_2 - Z_1) + iC(\hat{b}_2^+\hat{b}_2 - \hat{b}_1^+\hat{b}_1)]\hat{b}_2 + H.C.\}) \quad (59)$$

where H.C. denotes the Hermitian conjugate.

Due to conservation of energy, the sum of the output of the beam splitter ports (e and d) must be constant excluding quantum noise fluctuations. If the energy in the detector arm is increased because of positive interference, energy directed at the source must decrease (or vice versa). For the photon number operator $\hat{e}^\dagger\hat{e}$ in the second output port of the interferometer, one gets the same form of equation 59 but with a minus sign for the interference terms. The quantum noise is the fluctuation differences between each arm. In other words, we are looking at the difference of energy fluctuations between beam splitter outputs to the detector and source ports. This provides an experimental approach for studying quantum noise in an OCT system.

*5D. Other Vacuum Fluctuations Errors*

As already stated, quantum optics studies in OCT are limited. More work in controlling vacuum fluctuations needs to be done. We have already discussed vacuum fluctuations as a significant source of error entering the detector port of the beam splitter. But the beam splitter is not the only site in the interferometer where vacuum fluctuations influence OCT performance, but based on our current understanding it is likely the most relevant. We would just like to make several points about vacuum fluctuations in other parts of the system, generally influenced by the volume of the conduit. The first example is the influence of vacuum fluctuations in fiber versus free space OCT embodiments (where fluctuations are different). It is often envisioned that vacuum fluctuations in, for example, a closed space versus and open space, are the same [45,46]. But just examining the Casimir effect, as an example, illustrates why this is not the case. Here, if mirrors are placed facing each other in a vacuum, only certain frequencies of the vacuum can exist between them (nodes must exist at the interfaces). As the two mirrors move closer to each other, the longer waves will no longer 'fit'. Typically with OCT vacuum energy is contained in the 9 μm core optical fiber, which can be altered for example, by fiber bending. This leads to path distinguishability and noise.

In the second example of a vacuum influence outside the beam splitter, consider a detector that has a relatively low reflective surface versus one where the reflections are larger. The vacuum states entering the detector port interact with excited states of the field reflected from the detector. Therefore, the unbalanced energy fluctuations entering the beam splitter exit port are greater for the latter compared to the former. Path distinguishability is then greater. These represent two examples emphasize different system set-ups will (and often do) result in different vacuum fluctuations throughout the system, altering quantum noise levels [1,55].

*5E. Summary of Quantum Noise Reduction in OCT:*

Techniques to improve SNR through quantum noise reduction have been used successfully in other fields. However, with OCT work in this area is virtually non-existent. To advance OCT through quantum noise reduction, further experimental and theoretical work needs to be performed. This includes accounting for the broad bandwidth Gaussian field, the fact backreflections are coming simultaneously from different depths, and the difference in mass between the reference and sample arm targets.

## 6. Conclusions

Almost all current OCT theory is classical, but we argue that future advances lie in part with the quantum optics of OCT. We describe the need for a second quantization approach (rather than a semi-classical approach) to study the quantum mechanics of OCT. This paper focuses primarily on first order correlations, while part II will examine second order correlations and other quantum optics topics. The paper models the electric field in terms of a 'sea' of quantum harmonic oscillators, with the basis being Fock states. First orders correlations are described in terms of single photon interference through path indistinguishability. High intensity OCT is then build as a linear summation of these interferences in a Gaussian distribution. A comparison of the quantum and classical correlation function is used to illustrate classical limitations. The paper builds in part from the work of gravitational wave detection where extreme sensitivity are needed. However, OCT has distinct aspects that need to be accounted for including the broad bandwidth source, the sample arm target has multiple reflective surfaces, and smaller mirror sizes.

The direct application of these principles is examined in this paper with quantum noise reduction to improve signal to noise ratio. The major sources of quantum noise are vacuum fluctuations entering the detector port, photon counting error (PCE), and position probability amplitude uncertainty. The first two are derive quantitatively in the text. The position probability amplitude uncertainty, an important area in OCT ranging is described qualitatively but will be expanded upon in part II. Techniques for quantum noise reduction are discussed.

Though taking advantage of the quantum mechanical properties of light offers the opportunity for creating paradigm shifts in the field, little work is done in this area. By introducing the fundamentals of OCT quantum optics, describing the quantum mechanics of first order correlations, and examining techniques for quantum noise reduction as an example, this work seeks to advance investigations into this paradigm shifting area.

**Acknowledgements:** Dr. Brezinski's work is currently funded by National Institute of Health Grants R01 AR44812, R01 HL55686, R01 EB02638/HL63953, R01 AR46996 and R01 EB000419

**Figures**

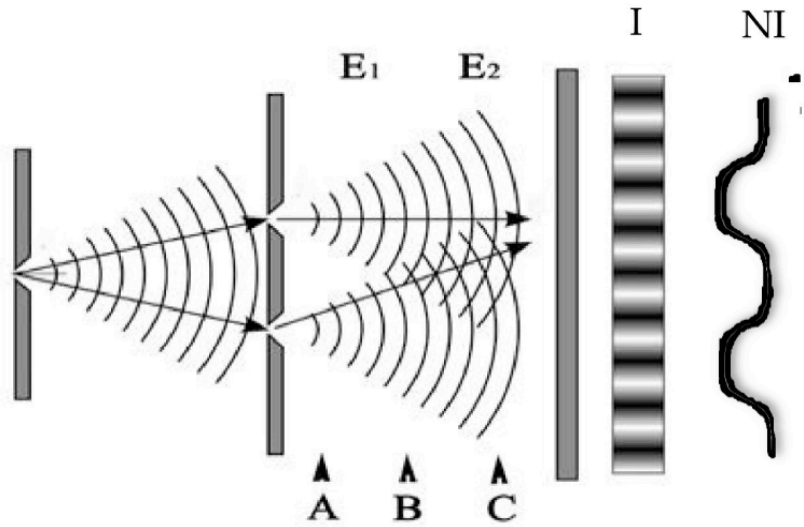

Figure 1: Illustration of Young's interferometry. In the figure, I represents interference, NI no interference, and E environmental interactions. A, B, and C are described in the text.

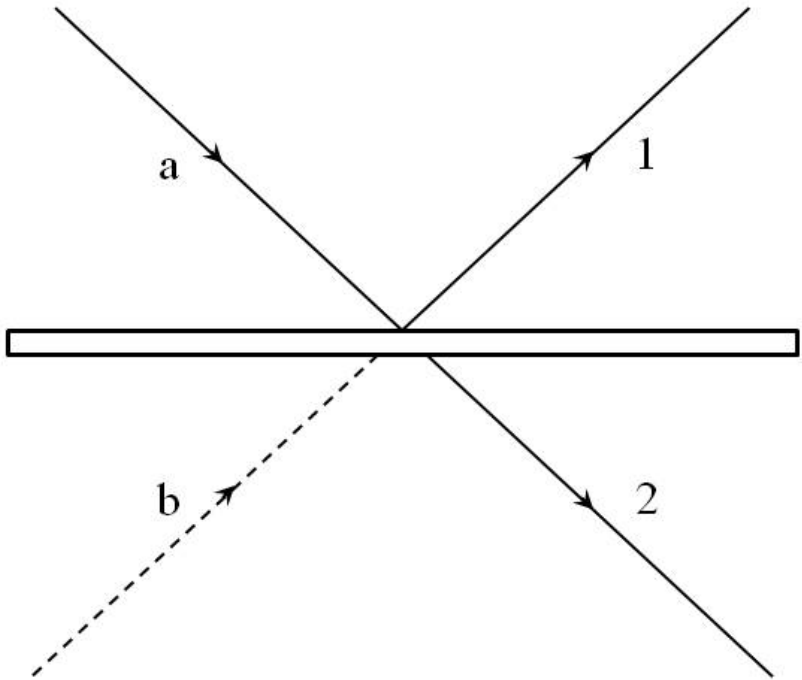

Figure 2: Illustration of the quantum beam splitter described in the text. One (1) is the source port and 2 is the detector/vacuum fluctuation ports.

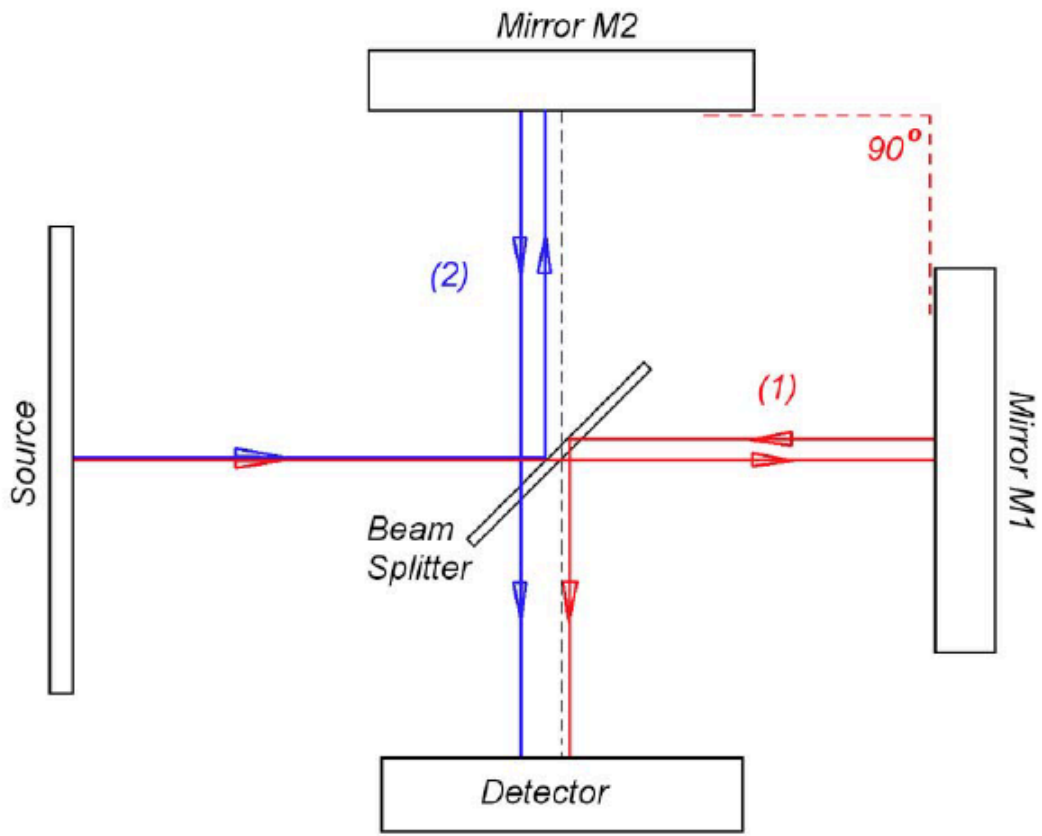

Figure 3: A Michelson interferometer is shown which is the base embodiment of OCT.

**Appendix: Measurement and Decoherence**

This appendix is for those who do not routinely work in quantum mechanics. It seeks to give an overview of the difference between decoherence and measurement. In the text we discuss decoherence and measurement (the latter popularly but inaccurately often termed 'collapse of the wavefunction'), two concepts that still are far from fully understood. In the Young's interferometer example, measurement is occurring at the screen while decoherence is occurring from reversible interactions with the environment (E). We operate under the description of reality that is non-local and where classical space-time is a manifestation of 'measurement' [3]. This is a concept we have described in detail elsewhere [64].

John von Neumann's version of measurement or 'collapse of the wavefucntion' has received the most attention for over the last three quarters of a century, so is presented here, where he viewed [18].

*Measurement as a transition from a quantum state into a classical state [1]:*
A quantum state:
- The deterministic, unitary, continuous time evolution of an isolated system that obeys Schrodinger's equation (or a relativistic quantum field theory or even more recent theories such as string theory). The state is generally a superposition of classical states until measurement. [More recent work would expand this to be non-local].

Measurement:
- The probabilistic, nonunitray, discontinuous change into a specific eigenvalue. In this view the 'observer' experiences a jump or collapse into a classical state.

In general, quantum systems exist in superpositions of basis states (eigenstates) which when not being measured or observed, evolve according to the time dependent Schrödinger equation or some equivalent evolution equation. However, when a measurement is taken, from an observer's perspective the state "leaps" or "jumps" to just one of the basis states and uniquely acquires a value of the property being measured. After the collapse, the system begins to evolve again according to the Schrödinger's equation or some equivalent time evolution equation.

Whether there is actually a discontinuity actually exists is a subject of considerable debate. Alternative approaches to this "measurement problem" than decoherence include Everett's relative interpretation and De Broglie-Bohm theory, but current opinion favors the decoherence based approaches (none of these approaches completely addresses the measurement process) [59,60]. Irrespective, measurement is an irreversible process where the system ends up in one of several eigenvalues. What state results is not deterministic and is dictated by probabilistic rules. The results of a single measurement can's be known a priore (with few exceptions) but the distribution of results of a large numbers of interactions can be known with extreme accuracy.

Decoherence can be thought of entanglement (reversibly or irreversibly) of the state (completely or in part) with the environment [4, 23-25]. This was illustrated with Young's experiment in figure 1. It is purely a quantum mechanical phenomenon. If it results in an irreversible loss of path indistinguishability at measurement, there is a loss of the original coherence. We will see how decoherence in part (but not completely) deals with the measurement problem.

The measurement problem (quantum to classical transition) generally has three components [25]:
1. The problem of preferred basis. An example is why are physical systems usually observed to be in definite positions rather than in superpositions of position.
2. The problem of non-observablity of interference.
3. The problem of outcomes. Why do measurements have outcomes at all and what selects out a particular eigenvalue among the different possibilities described by the quantum probability distribution.

Decoherence seems to address 1 and 2. With respect to one, decoherence induced pointer states (environmentally selected states for examples), select out a preferred basis. With respect to two, loss of interference occurs when indistinguishable paths become distinguishable via environmental interactions (described in text). In general for decoherence to occur, the environment that the state is entangled to is different for each potential path, environmental elements do not significantly have correlations with each other, and rapidly dissipates the interaction to other environmental elements (large degree of freedom).

However, decoherence does not appear to provide an answer to the third component of measurement. Measurement results in specific eigenstates that occur with quantum mechanical probability distributions [3]. To date, a generally accepted theory how (or if) decoherence can select out specific eigenvalues with specific probabilities of occurring has yet to be accepted.